\documentclass[12pt]{article}
\usepackage{amssymb} 
\usepackage{amsfonts}
\usepackage{amscd}
\newcommand{\beq}{\begin{equation}}
\newcommand{\eeq}{\end{equation}}
\newcommand{\beqs}{\begin{eqnarray}}
\newcommand{\eeqs}{\end{eqnarray}}

\newcommand{\tr}{{\mathrm{tr}}}

\newcommand{\rank}{{\mathrm{rank}}}
\newcommand{\Dhyper}{{\mathcal D}}

\newcommand{\Lie}{{\mathrm {Lie}}}
\newcommand{\ad}{{\mathrm {ad}}}
\newcommand{\Ad}{{\mathrm {Ad}}}

\newcommand{\dd}{{\mathrm d}}
\newcommand{\ee}{{\mathrm e}}

\newcommand{\ZZ}{{\mathbb Z}}

\newcommand{\CC}{{\mathbb C}}

\newcommand{\cartan}{{\bf h}}
\newcommand{\lie}{{\bf g}}
\newcommand{\torus}{{\bf t}}
\newcommand{\Cech}{{\v{C}ech}}

\begin{document}

\begin{titlepage}
\begin{flushleft}
       \hfill                       SISSA 125/99/EP\\
       \hfill                      {\tt hep-th/yymmxxx}\\
\end{flushleft}
\vspace*{3mm}
\begin{center}
{\LARGE Non-Abelian Gerbes from Strings on a Branched Space-Time\\}
\vspace*{12mm}
{\large Jussi Kalkkinen \footnote{\tt kalkkine@sissa.it}}\\
\vspace*{4mm}
{\it SISSA, via Beirut 4, Trieste 34014, Italy\\}
{\it and\\}
{\it INFN, Sezione di Trieste\\}
\vspace*{10mm}
\end{center}

\begin{abstract}
As superstring solitons that carry Neuveu--Schwarz charge
can be described in terms of gerbes, one expects 
 non-Abelian gerbes to appear
e.g.~in the exotic six-dimensional world-volume theories of 
coinciding NS5 branes.
We consider open bosonic 
strings on a space-time that is branched in such a way that
the $B$-field is provided 
with the same Lie algebra structure as the world-volume 
gauge field on a D-brane. 
These considerations motivate a generalization
of the cocycle conditions and the transformation rules of an Abelian gerbe
in hypercohomology. 
The resulting system incorporates in a natural way
the NS two-form, the RR gauge field, the Chan--Paton gauge field, the
relevant gauge transformations and the holonomies associated to
Wilson surface observables. 
\end{abstract}
\end{titlepage}

\tableofcontents
\pagebreak

\section{Introduction}

One of the most intriguing systems in string theory
is that of $r$ overlapping NS5 branes 
\cite{Strominger:1996ac} 
in the decoupling limit.
The world-volume theory of an isolated NS5 brane in type IIA 
is given by a $N=(0,2)$ supersymmetric theory in six-dimension, and
it therefore involves self-dual antisymmetric tensorfields 
\cite{Romans:1986er, Riccioni:1997np}.
For several NS5-branes the dominant low energy degrees of freedom
are tensionless strings that arise in M-theory from M2-branes 
suspended between M5-branes when the M5-branes' world-volumes
coincide. These theories appear also in type IIB compactified
on $K3$ \cite{Witten:1995zh,Witten:1996em}, in which formulation
it becomes clear that the appearing tensionless strings 
are not fundamental strings and that they have  $ADE$ 
gauge symmetry. The heterotic description 
involves a small $E_8$ instanton
\cite{Ganor:1996mu} at the core of which gauge symmetry is enhanced.
Hence, it is natural to look for a way to describe these systems
in terms of a local quantum field theory that involves a 
non-Abelian self-dual two-form in six dimensions. 
However, such a theory does not exist \cite{Bekaert:1999dp} (cf.~also \cite{Witten:1996hc}),
and one needs an indirect or nonperturbative description. 
Wilson surfaces and loop equations 
in these systems have been studied in 
\cite{Ganor:1997nf}. There is also a M(atrix) theory construction 
\cite{Seiberg:1997zk}.  

Another interesting system involves the $N=(1,1)$ `new' gauge theories of Witten
in six-dimensions \cite{Witten:1998kz}. 
In type IIB they appear on a $\CC^2/\ZZ_r$
orbifold when  nonperturbative closed string states appear
from open strings that start and end on points in $\CC^2$
that are identified by the $\ZZ_r$
action. In low energies the dynamics should however reduce to
the  six-dimensional infrared-free $N=(1,1)$ Yang--Mills theory.

The bulk origin of the anti-symmetric tensor field is the 
Neuveu--Schwarz two-form $B$, the gauge field of fundamental
string charge. There are many interesting phenomena connected
to it, such as the appearance of noncommutative Yang--Mills 
theories in a constant background condensate 
\cite{Douglas:1998fm,Schomerus:1999ug,Seiberg:1999vs}. If the curvature $[H]$ 
of the $B$ field is  a torsion class in integral cohomology, 
D-brane charges can be  classified \cite{Minasian:1997mm}
in a twisted version of K-theory \cite{Witten:1998cd}, and 
the Chan--Paton gauge fields appear as connections on a module of 
a noncommutative algebra \cite{Kapustin:1999di}. 
If the curvature is not a torsion class
then classification  in terms of K-theory fails. 
For general
curvature $H$, and thus bound states that involve nontrivial
NS five-brane charge, the classification problem is still open.

The proper mathematical framework for treating these
three-form fluxes seems to be that of gerbes \cite{Giraud}. 
In all generality gerbes are sheafs of grupoids 
\cite{Brylinski,Finlay}, but they can be understood 
more concretely as collections of local principal bundles 
and their isomorphisms. As such they also include the modules 
of noncommutative spaces \cite{Connes} that appear in noncommutative
Yang--Mills. Abelian gerbes allow for a geometric
interpretation in terms of local line bundles
\cite{Chatterjee,Hitchin}, and of  hypercohomology 
\cite{Brylinski,Alvarez:1985es,Gawedzki:1987ak}.
The role of hypercohomology
is to provide a differential geometric framework for
studying gerbes. Physically this corresponds to
finding the correct local degrees of freedom for field theory. 
However, this has been done until now only for Abelian gerbes. 
Non-Abelian gerbes do exist, actually the concept was 
originally introduced to formulate noncommutative 
cohomology. However, the description uses holonomies and isomorphisms,
which physically corresponds to a Wilson loop, or surface, observables. 

Recently it was shown \cite{Kalkkinen:1999uz} that the local line 
bundles of Hitchin \cite{Hitchin} 
indeed appear in effective type IIA solutions in massive 
supergravity, when NS5 branes and D6-branes are involved. 
The same considerations also gave reason to suspect 
that gerbes should enter whenever NS charges are involved, including 
the world-sheet theories on D-branes. 
However, these theories involve non-Abelian bundles on the world-volumes, and
the Abelian gerbes are clearly not suited for describing them.
In this article our aim is to 
find a straight forward non-Abelian 
generalization of the Abelian hypercohomology underlying a general 
gerbe. 

In order to do this 
we shall consider the quantum theory of
strings on a branched space-time or, more concretely, 
multivalued $B$-fields.
These models are adequate for describing strings both on an
orbifold $\CC^2/\ZZ_r$ in type IIB description, and on 
$r$ coinciding NS5 branes, when each one of them 
is carrying an independent $B$-field. 
This serves as a rough bosonic model for strings in both six dimensional
$N=(0,2)$ and $N=(1,1)$ systems as we are not imposing
self-duality constraints on the two-form. In fact, what follows does not
depend on the dimensionality of the brane, either.   
These arguments are
sufficient for establishing 
what fields and which symmetries to expect in a differential 
geometry description of gerbe. 
A simple  non-Abelian 
generalization of the cocycle conditions and symmetry transformations of 
an Abelian gerbe then yields a strongly constrained system which 
fits well in the physical picture; it involves a collection of essentially Abelian
RR fields, and a Chan--Paton and NS two-form, which will turn out to be 
non-Abelian, though  in a somewhat 
restricted sense. The result is suitably Abelian as to allow
us to use some field theory intuition in these systems, but display
the underlying non-Abelian structure clearly enough, so that we see
why and when we would have to move from a local
field theory description to a nonperturbative one, and to Wilson surfaces.

The plan of the paper is as follows: In the next section we shall consider 
string sigma-models and non-Abelian currents on a branched space-time. 
These results motivate in Section 3 a non-Abelian generalization of 
Abelian hypercohomology. In Section 4 we  show
how the mathematical framework fits together with what we know about
superstring solitons, and their effective low energy theories.

\section{World-sheet actions and currents}

We shall start by considering  the bosonic string sigma-model 
in a framework that naturally accommodates what we know about 
open string dynamics in the presence of several D-branes.

Consider D-branes $Q_i$, where $i=1,\ldots,r$. Each of 
these branes carries a Chan--Paton vector potential $A_{M,i}$, where 
$M,N=1,\dots,D$ are space-time indices.
In the sigma-model this condensate field is integrated over
the components of the boundary of the world-sheet $\Sigma$ on 
different D-branes, namely $\partial \Sigma_i$.
It is natural to think of the boundary as a vector in $H^1(M) \otimes 
{\cartan}$ where ${\cartan}$ is a vector space of dimension $r$ with basis
 $\{{\bf e}^i \}$. 
Similarly, $A$ should be thought of as an element of $\Omega^1(M,{\cartan}^*)$. 
One particular realization of this is to take ${\cartan}$ to be 
the Cartan subalgebra of some Lie algebra $\lie$ of rank $r$.

Let us in particular consider the configuration where all 
of the D-branes $Q_i$ lie on top of each other, and the boundaries 
$\partial\Sigma_i$ coincide with $\partial\hat\Sigma$. Then we can write
\beqs
\int_{\partial\Sigma} A = \sum_i \int_{\partial\Sigma_i} A_{M,i}~
\partial_{\varphi}X^{M}_i ~\dd \tau = \int_{\partial\hat\Sigma} ~(A_{M},~ 
\partial_{\varphi}X^{M}) ~\dd \tau~.
\eeqs
where $\varphi$ denotes $\tau$ for parallel and $\sigma$ 
for transverse coordinates to the D-brane,
and $(~,~)$ is the Killing form of the Lie algebra. 
The index of the boundary component became in this way formally
an index of the at the moment diagonal {coordinate matrix} $X^M$.

Let us next turn to the B-field, and the full world-sheet. 
Consider, for simplicity, a world-sheet  that is composed of disjoint 
{cylinders} $\Sigma_{ij}$ that connect 
the boundaries
$\partial\Sigma_i$ and $\partial\Sigma_j$. As we already attached the vectors
${\bf e}^i$ and ${\bf e}^j$ to them 
it is natural to attach their difference ${\bf e}^i - {\bf e}^j = \underline\alpha$
to the interpolating world-sheet $\Sigma_{\underline\alpha}$, 
hence $\partial \Sigma_{\underline\alpha} = 
\partial\Sigma_i - \partial\Sigma_j = {\bf e}^i - {\bf e}^j = \underline\alpha$.
In particular in the case that the Lie group $\lie$ is just  $A_r$ 
the vector $\underline\alpha$ is one of its  roots.

In string theory there is only one bulk $B$-field. However, 
it is natural to associate different pull-backs of this field 
$B_{MN,\underline\alpha}$ for each component of the world-sheet 
$\Sigma_{\underline\alpha}$. 
For disjoint cylinders we can write without any loss of generality
\beqs
B_{MN,\underline\alpha} = (B_{MN}, ~\underline\alpha^{\vee}) ~,
\label{Broots}
\eeqs
where $\underline\alpha^{\vee}$ is the coroot\footnote{We follow the 
conventions of \cite{Fuchs}.}. 
The pertinent world-sheet integrals can now be written in the form
\beqs
\int_{\Sigma} B = \sum_{\underline\alpha} 
\int_{\Sigma_{\underline\alpha}} (B, ~\underline\alpha^{\vee})~.
\eeqs

\subsection{Non-Abelian currents}
\label{current}

Until now all Lie algebra has been used for
keeping track of disconnected components of the world-sheet.
Consider now a configuration where the cylinders fuse  
into one geometrical object. Let us take this limit 
in such a way that the components of the 
fields $B_{MN}$ and $A_M$ stay independent on each component cylinder;
This means that these fields effectively live on different branches of 
space-time\footnote{This approach can be naturally formulated in Connes' noncommutative 
geometry \cite{Connes}. It has been studied in open string context for instance in
\cite{Kalkkinen:1997ci}.}. For the
vector fields this is the standard limit of coinciding D-branes.
As to what concerns the $B_{MN}$  field, this kind of a situation could arise
for instance at an orbifold such as $\CC^2/\ZZ_r$ where
the different components come  
from different fundamental domains of the $\ZZ_r$ action,
though we do not have an explicit CFT construction for this.

Suppose further that we have two cylinders $\Sigma_{\underline\alpha}$ 
and $\Sigma_{\underline\beta}$ that overlap on one of their boundaries. 
In the  limit we are taking both of the cylinders are forced to occupy 
the same part of space-time, and as they are connected, one is simply folding one 
on top of the other. However, as far as the $B$ field is concerned 
we could have equally well  
started with the combined cylinder $\Sigma_{\underline\alpha + \underline\beta}$. 
This means that even in the limit where one allows the $B$ field be independent 
on different world-sheets we have to impose a consistency condition
\beqs
B_{\underline\alpha} + B_{\underline\beta} = B_{\underline\alpha + \underline\beta}~.
\eeqs
Assuming (\ref{Broots}) solves this condition, as anticipated.

The result of this analysis is hence that the string world-sheets 
on different branches of space-time and the $r$ different boundaries of the 
cylinders can both be 
associated to the same Cartan subalgebra of a Lie algebra $\lie = A_r$, 
and the connecting cylinders
to the roots of the same algebra. 
This argument generalizes to world-sheets with an arbitrary number of boundary components. 
For instance, given the boundary 
$\Sigma = {\bf e}^1 + {\bf e}^2 - {\bf e}^3 = {\bf v}$, the correct 
$B$-field  proportional to $(B,{\bf v})$. These world-sheets belong 
naturally to some representation of $\lie$, with Dynkin labels  ${\bf v}$. 
Note also that we are not restricted to the unitary series, 
but modding by a suitable symmetry we get all of the 
simply laced Lie algebrae in the $ADE$ series. 

In order to learn how to describe these models in effective field theory we 
need a theory that can accommodate non-Abelian $B$ fields. 
This question will occupy us for the rest of the paper.
Define\footnote{$H^i$ are Cartan generators,  
$\underline\alpha$ the positive roots, and $E^{\underline\alpha}$ their 
generators \cite{Fuchs}.}
$A_M = A_{M,i}~H^i$ and 
$B_{MN} = B_{MN,i}~H^i$. 
It is also useful to introduce  the non-Abelian line element
\beqs
\dd X^M &=& \dd \tau \partial_{\varphi} X^M_i H^i + \dd z 
\partial X^M_{\underline\alpha} E^{\underline\alpha}   \label{line-ele}
 + \dd \bar{z} \bar{\partial} X^M_{-\underline\alpha} E^{-\underline\alpha}~.
\eeqs
Then the full non-Abelian world-sheet action can be succinctly 
summarized in
\beqs
S &=& \tr \int_{\hat\Sigma} (G_{MN} + B_{MN})~ \dd X^M \dd X^N + \tr 
\int_{\partial\hat\Sigma} A_M ~\dd X^M \label{ncgmodel}~.
\eeqs
We stress that the Lie algebra indices $i$ for a boundary component and 
$\underline\alpha$ for a connecting world-sheet 
arose geometrically when one evaluated  
coordinate functions on different components of the world-sheet.

From open string interactions between $r$ 
coinciding  D-branes we know \cite{Polchinski:1995mt,Witten:1996im}
that the gauge fields $A_M$ 
can be extended to the full 
Lie algebra $\lie$, and that the scalars that appear
as transversal coordinates take values in the same space. 
In (\ref{ncgmodel}) this means that we should allow $A_M$ and hence $\partial_{\varphi} X^M$
take arbitrary values in the full Lie algebra, and interpret 
\beqs
\int_{\partial\hat\Sigma} \delta(x - x(\tau))    ~\dd X^M
\eeqs
as the non-Abelian current\footnote{Note that $X^M$ 
denotes a sigma-model coordinate in all of the space-time 
directions; The physical transverse coordinates of the brane 
are included in $A_M$.} carried by a particle moving along $\partial\hat\Sigma$.

Extending this procedure to the bulk fields (on the world-sheet) 
$B_{MN}$ and $X^M(z,\bar z)$ is tempting. This would formally
mean that we introduce new degrees of freedom to the theory, 
namely the non-diagonal components of the $B$-field. These components 
couple to coordinate functions $X^M_{\underline\alpha}$ on {\em different} 
world-sheets and would seem to correspond to strings propagating from one 
world-sheet -- or sheet of space-time -- to another.

\subsection{Effective actions and symmetries}

The action (\ref{ncgmodel}) describes the
coupling of a macroscopic string $\hat\Sigma$
to the string condensates. From
the effective field theory point of view it hence appears 
as a non-Abelian current. In order to address the dynamics of
the full background fields $A$ and $B$ one produces the generating functional
of their interactions from the path integral evaluated in 
the presence of this current.
In the absence of the $B$ field the functional is just 
the Wilson line \cite{Tseytlin:1988ww}
\beqs
\ee^{-F[A_c, B_c]} &=& \Big\langle ~\tr ~{\rm Pexp} - i 
\int_{\partial\hat\Sigma} A_c \Big\rangle~. \label{wilson}
\eeqs
It can be evaluated assuming that one is allowed 
to neglect derivative terms and 
commutators of the field strength, and the result is Tseytlin's 
generalization of the DBI action \cite{Tseytlin:1997cs}.

This argument also tells us how to study {non-Abelian} 
$B$-fields in string theory. We like to simply insert the 
current (\ref{ncgmodel}), and ask what the resulting 
generating function tells us about the dynamics of the field. 
As the microscopic description for non-Abelian $B$ field is 
lacking we have to rely on indirect arguments, such as those that 
make use of the Wilson line above, or general underlying 
structures associated to the field, gerbes.

\label{gaugesymm}
In order to find out what gauge symmetries we have, 
let us in particular assume first $B=\dd C$, where $C$ is a diagonal, 
but make no restrictions on $A$.
Then we can eliminate the $B$ field, and confirm that 
the path integral (\ref{wilson}) is invariant under the transformations  
$A+C \longrightarrow k^{-1}(A + C + \dd)k$. 

The coordinate system in which  $B$ is diagonal in 
isospin indices tells us what the geometrical 
direction in the isospin space should be -- much like the coordinate 
system in which the gauge field of a D-particle 
is diagonal defines what we mean by asymptotic space-time. 
However, when $A$ is generally non-Abelian, there should 
not be a particularly preferred choice of this diagonal 
direction because we can always change the 
basis in  (\ref{ncgmodel}). Hence, we may have an 
independent freedom to change $B$ by an isospin rotation. 

In all, we have found three, 
as it seems, independent symmetries of the theory
\begin{itemize}
\item[{\bf (G1)}]
The generalized NS symmetry for a local one-form $\eta \in \Omega^1(Q,\cartan)$
where $\cartan \subset \lie$ is the Cartan subalgebra
\beqs
A  & \longrightarrow & A - \eta \\
B & \longrightarrow & B + \dd \eta ~.
\eeqs
This symmetry relies heavily on the fact that $\eta$ is assumed diagonal
with respect to the basis (\ref{line-ele}).
\item[{\bf (G2)}]
The ordinary non-Abelian gauge symmetry $k:Q \longrightarrow G$
\beqs
A  & \longrightarrow & k^{-1} (A + \dd) k \\
B & \longrightarrow & B  ~.
\eeqs
\item[{\bf (G3)}]
Provisionally, we include also the 
choice of the physical direction in isospin space  
$h:Q \longrightarrow G$
\beqs
A  & \longrightarrow & A \\
B & \longrightarrow & h^{-1} B h~. 
\eeqs
Note, however, that the non-Abelian DBI action is not invariant under this 
symmetry unless $k=h$.
\end{itemize}
Next we shall try to combine these local symmetries and fields 
in a global framework. For this we shall, however, have to 
find out how to describe a non-Abelian gerbe in
terms of differential geometry.

\section{Hypercohomology}

Let us consider a space-time manifold $X$ and a fixed\footnote{Neither the patches in the cover 
nor their intersections need to be contractible. \Cech-cohomology
does not depend on the cover, if the cover is fine enough. Here we shall not
dwell on the dependence of the construction on the choice of cover. See, however, end of Section 
\ref{rajotus}.} 
open cover  $\{ {\cal U}_{\alpha}\}$.
The isomorphism class of 
an Abelian one-gerbe with connective structure and curving
is given by a two-cocycle in the hypercohomology of the complex $C^{\infty} 
\longrightarrow \Omega^1 \longrightarrow \Omega^2$ 
\cite{Alvarez:1985es,Gawedzki:1987ak}. 
A gerbe can hence  be thought of as  a two-cocycle in the  
hypercohomology ${\cal H}^2$ of 
\Cech-cocycles with de~Rham forms on the coefficient sheaves 
\cite{Brylinski}. 
A representative of this class is then a closed 
two-cochain\footnote{The index in square brackets denotes the de Rham form-degree, and 
Greek alphabet is used to label the intersections of local patches
where the object is defined. Thus, for instance, $A^{[1]}_{\alpha\beta}$ 
is a one-form defined on every twofold intersection of local charts, 
namely ${\cal U}_{\alpha} \cap  {\cal U}_{\beta} = {\cal U}_{\alpha\beta}$.} 
\beqs
\underline{w} &=& \Big[ ~ g^{[0]}_{\alpha\beta\gamma}, 
~ A^{[1]}_{\alpha\beta}, ~B^{[2]}_\alpha ~ \Big ] \label{gerbe}~, 
\eeqs
and any two representatives of the class are connected by  a shift 
with an exact term, which will just turn out to be a
gauge transformation. We shall give the cocycle conditions and the gauge 
transformation rules below.

The \Cech-coboundary operator $\delta$ acts by adding an index to 
$h_{\alpha\beta}$, such that for instance 
\beqs
\delta h_{\alpha\beta\gamma} = h_{\beta\gamma} ~h^{-1}_{\alpha\gamma} 
~h_{\alpha\beta}~.
\eeqs
The \Cech-indices always remain antisymmetric, and $\delta^2 = 0$. 
The zero-forms are multiplicative and the higher de Rham forms additive.
The coboundary operator of the complex introduced above  
is, when acting on an $i$-form,
\beqs
\Dhyper =  \dd_{{\rm de Rham}} + (-1)^i \delta_{\mbox{\scriptsize \Cech}} ~.
\eeqs
The statement that $\underline{w}$  be closed under this coboundary operator
$\Dhyper \underline{w} = 0$ gives  the cocycle conditions
\beqs
g_{\beta\gamma\delta} ~g^{-1}_{\alpha\gamma\delta} 
~g_{\alpha\beta\delta} ~g^{-1}_{\alpha\beta\gamma} = 1 
&  \mbox{ on }  & {\cal U}_{\alpha\beta\gamma\delta}  \label{co1} ~,\\
g^{-1}_{\alpha\beta\gamma} ~\dd g_{\alpha\beta\gamma} - 
A_{\beta\gamma} + A_{\alpha\gamma} - A_{\alpha\beta} = 0 
&  \mbox{ on }  & 
{\cal U}_{\alpha\beta\gamma} \label{co2}  ~, \\
\dd A_{\alpha\beta}+ B_{\beta} - B_{\alpha}
 = 0 &  \mbox{ on }  & 
{\cal U}_{\alpha\beta}\label{co3} ~.
\eeqs

Gauge symmetry arises from shifting the gerbe $\underline{w}$ by
an exact term 
\beqs
\underline{w}' &=& \underline{w} + \Dhyper \underline{\lambda}~,
\eeqs
where $\underline{\lambda}$ is a cochain in the lower complex  
$C^{\infty} \longrightarrow \Omega^1$
\beqs
 \Big[ ~ h^{[0]}_{\alpha\beta}, ~ \eta^{[1]}_{\alpha}~ \Big]~.
\eeqs
This makes $g_{\alpha\beta\gamma}$ into an ordinary 
\Cech-cocycle, $A_{\alpha\beta}$ into a connection of a 
line bundle defined on ${\cal U}_{\alpha\beta}$, and 
$H = \dd B_{\alpha}$ into a globally 
defined three-form. Concretely,
\beqs
g_{\alpha\beta\gamma}' &=& g_{\alpha\beta\gamma} h_{\beta\gamma} h^{-1}_{\alpha\gamma} 
h_{\alpha\beta}   \label{ga1} \\
A_{\alpha\beta}' &=& A_{\alpha\beta} + h^{-1}_{\alpha\beta} \dd h_{\alpha\beta}  
\label{ga2} - \eta_{\beta} + \eta_{\alpha}  \\
B_{\alpha}' &=& B_{\alpha} + \dd \eta_{\alpha} \label{ga3}~.
\eeqs

Similarly, $\underline{\lambda}$ becomes a one-cocycle, if one imposes the condition 
$\Dhyper \underline{\lambda}=0$. It is easy to see that $\underline{\lambda}$ 
is then a principal 
bundle with connection $\eta$ and transition functions $h$.

\subsection{Non-Abelian gerbes}
\label{fullgen}

A straight forward attempt to simply add a Lie algebra 
(or group) index to $\underline{w}$
fails as it seems to be rather difficult 
to define a non-Abelian  generalization of the 
cocycle condition (\ref{co1}). Indeed, there is 
no non-Abelian \Cech-theory, and hence
a non-Abelian hypercohomology formulation is absent as well.
Gerbes were originally invented for studying
non-Abelian cohomology \cite{Giraud}, and they have appeared e.g.~in \cite{Finlay}
as Dixmier--Duady sheaves of grupoids \cite{Brylinski,Finlay}.  
Abelian gerbes carry a natural differential geometry,
but this structure has not been extended to the non-Abelian case 
\cite{Brylinski,Hitchin}. For physics the differential complexes
are rather essential as they correspond to physical fields. 
The formulation in terms of
grupoids corresponds rather to a QFT formulation in terms 
of Wilson line and other holonomy operators 
\cite{Freed:1999vc}.

The idea of the present construction comes from D-particle physics. 
There a number of classical particles moving in the target space is
described by a diagonal matrix. In physical processes this coordinate
matrix has to be asymptotically diagonal -- in some basis -- but the 
dynamics of the theory allow processes where also off-diagonal elements 
of the coordinate matrices are excited. Then the notion of local space-time 
vanishes and we are in the realm of noncommutative geometry, or 
stringy geometry.
In particular, one could consider a process where the in-coming and out-going
particles live in different Cartan subalgebrae of the pertinent group; 
then the 
definition of space-time seems to have changed under the process.

In what follows we shall study a non-Abelian generalization of the Abelian 
hypercohomology relevant for gerbes, which is in many respects still 
essentially Abelian. One circumvents the problem of defining non-Abelian 
\Cech-cohomology  
in formulae (\ref{co1}) and  (\ref{ga1}) by assuming that the system is actually  
Abelian on threefold intersections. One can still
allow non-Abelian behaviour outside 
these by assumption isolated patches on the manifold.
More concretely we choose for each threefold 
intersection ${\cal U}_{\alpha\beta\gamma}$ a 
fixed torus $T_{\alpha\beta\gamma}$ inside a Lie group $G$, and 
assume that the \Cech{} two-cocycles $g_{\alpha\beta\gamma}$
as well as the restrictions of the gauge transformations $h_{\alpha\beta}$ 
on any ${\cal U}_{\alpha\beta\gamma}$
take values in the same fixed torus. 
Outside triple intersections we need not constrain these 
transformations. However, we shall have to impose further
assumptions on other fields.

The obvious generalization of the one-form is now to make it  
a connection on a principal 
$G$-bundle on ${\cal U}_{\alpha\beta}$. 
The cocycle condition (\ref{co2}) can then be 
accepted as is; for the fixed torus part the condition is 
then just a collection of Abelian equations, for the rest of
the Lie algebra it reduces to a cocycle condition that does not involve 
$g_{\alpha\beta\gamma}$. This restriction on the form of the one-form 
$A_{\alpha\beta}$ only constrains 
the field on the triple intersections. 
Under the non-Abelian gauge transformations we have to impose 
further  
\beqs
\delta(\Ad(h_{\alpha\beta}) A_{\alpha\beta}) = 0~. \label{A-3}
\eeqs
This condition can be solved by assuming that on 
three-fold intersections the  one-forms take values in the Lie algebra of the fixed torus 
$\torus_{\alpha\beta\gamma} = \Lie(T_{\alpha\beta\gamma})$.
Finally, the two-form field should be made Lie algebra valued as well.

We are now ready to write down an ansatz for 
the non-Abelian generalization of hypercohomology.
Our strategy will be to try to find  a non-Abelian generalization
for the two-cochain $\underline{w}$, together with the corresponding
new cocycle conditions and transformation rules. In addition to this 
two-cocycle it will turn out that it is actually necessary to also include 
a fixed one-cochain $\underline{v}$ together with 
its transformation rules. This 
one-cochain need not be closed.
 In the following, we consider then a fixed two-cocycle and 
a fixed one-cochain
\beqs
\underline{w} &=& \Big[ ~ g^{[0]}_{\alpha\beta\gamma}, 
~ A^{[1]}_{\alpha\beta}, ~B^{[2]}_\alpha ~ \Big ] \\
 \underline{v} &=& \Big[ ~ \phi^{[0]}_{\alpha\beta}, 
~ \chi^{[1]}_{\alpha} ~ \Big ]~,
\eeqs
and the action of the two cochains 
\beqs
 \underline{\lambda} &=& \Big[ ~ h^{[0]}_{\alpha\beta}, 
~ \eta^{[1]}_{\alpha} ~ \Big ] \\
 \underline{\kappa} &=& \Big[ ~ k^{[0]}_{\alpha} ~ \Big ] 
\eeqs
on them. Here all of the one and two-forms take values in the Lie algebra 
$\lie$, and the functions in the corresponding Lie group. 
As it will turn out, $\underline{v}$ describes a local 
principal bundle on each coordinate patch, and isomorphisms 
between bundles on different though intersecting patches. 
If the bundle were global, the cocycle conditions
\beqs
\phi_{\beta\gamma} \phi_{\gamma\alpha} \phi_{\alpha\beta}  &=& 1\\
\phi^{-1}_{\alpha\beta} (\chi_\alpha + \dd)  \phi_{\alpha\beta} - \chi_\beta &=& 0
\eeqs
would be satisfied everywhere. In the Abelian case the cocycle conditions can be succinctly
stated by saying that $\underline{v}$ is closed, $\Dhyper\underline{v}=0$.
The bundles  $\underline{v}$ could fail to
be a global bundle if the class $g_{\alpha\beta\gamma} = \phi_{\beta\gamma} 
\phi_{\gamma\alpha} \phi_{\alpha\beta}$ were not trivial. 
With this identification  $\underline{w}$ actually measures to what extent 
the  structure $\underline{v}$ is local; it is hence 
an obstruction  for making  $\underline{v}$ a global bundle.

In this nested structure $\underline{\kappa}$ acts as gauge transformations
on $\underline{\lambda}$ and $\underline{v}$, and both $\underline{\kappa}$ and 
$\underline{\lambda}$ act on  $\underline{w}$. 
In particular, the action of  $\underline{\lambda}$ on $\underline{w}$ is 
\beqs
g_{\alpha\beta\gamma}' &=& g_{\alpha\beta\gamma} h_{\beta\gamma} h^{-1}_{\alpha\gamma} 
h_{\alpha\beta}  ~, \\
A_{\alpha\beta}' &=&  h^{-1}_{\alpha\beta} (A_{\alpha\beta}  - \eta_{\beta} + \eta_{\alpha} 
+  \dd) h_{\alpha\beta}  ~,  \\
B_{\alpha}' &=& B_{\alpha} + F(\eta_{\alpha})~, \label{B-eta} 
\eeqs
where $F(x) = \dd x + x \wedge x$.  
Having $\underline{v}$ at our disposal we could have 
defined $B_{\alpha}' = B_{\alpha} + D(\chi_{\alpha})\eta$ 
instead of (\ref{B-eta}), but this would lead to
wrong transformation properties in (\ref{Hdef}) later on.
The gauge transformations 
$\underline{\kappa}$ act according to 
\beqs
h_{\alpha\beta}' &=& k^{-1}_\alpha h_{\alpha\beta} k_\beta \\
\eta'_\alpha &=& k^{-1}_\alpha (\eta_\alpha + \dd) k_\alpha~. \label{ggaa1}
\eeqs
on the local principal bundles. The action on $\underline{w}$ is both through
\beqs
B_{\alpha}' &=& k_\alpha^{-1} B_{\alpha} k_\alpha~, \label{B-k}
\eeqs
and through the action induced through $\underline{\lambda}$ in (\ref{ggaa1}).

The highest object obeys the cocycle condition 
$\Dhyper \underline{w} = 0$ namely,
\beqs
g_{\beta\gamma\delta} ~g^{-1}_{\alpha\gamma\delta} 
~g_{\alpha\beta\delta} ~g^{-1}_{\alpha\beta\gamma} = 1 
&  \mbox{ on }  & {\cal U}_{\alpha\beta\gamma\delta}  ~,\\
g^{-1}_{\alpha\beta\gamma} ~\dd g_{\alpha\beta\gamma} - 
A_{\beta\gamma} + A_{\alpha\gamma} - A_{\alpha\beta} = 0 
&  \mbox{ on }  & 
{\cal U}_{\alpha\beta\gamma}  ~,\\
F(A_{\alpha\beta}) + B_{\beta} - B_{\alpha}
 = 0 &  \mbox{ on }  & 
{\cal U}_{\alpha\beta} \label{B-A} ~.
\eeqs
We call the  collection of fields $\underline{w}$ a non-Abelian one-gerbe 
if it satisfies these consistency conditions.

Where ever the ``zero-gerbes'' $\underline{v}$ and $\underline{\lambda}$ 
obey the cocycle conditions  $\Dhyper \underline{v} =0$ or $\Dhyper \underline{\lambda} =0$ 
they are actually locally defined principal $G$-bundles. 
The former should not be assumed globally closed $\Dhyper 
\underline{v} \neq 0$,
as otherwise it would indeed extend to a global principal bundle, and the 
obstruction to this $\underline{w}$, in which we are actually interested, should vanish. 
Also assuming  $\underline{\lambda}$ closed 
would imply that it act at least in the Abelian case trivially on 
$\underline{w}$. Also $\underline{\kappa}$ has a geometrical interpretation: 
it is just the set of gauge transformations  of $\underline{\lambda}$. 

We should take care that the cocycle conditions  $\Dhyper \underline{w} =0$ 
are invariant under the action of $\underline{\kappa}$ and 
$\underline{\lambda}$. The first two conditions are still trivial, 
thanks to the assumption that all relevant fields collapse to tori 
on triple intersections, cf.~(\ref{A-3}). 
The last cocycle condition, however, gives a 
restriction on
$\underline{\lambda}$ and $\underline{\kappa}$. In all generality
\beqs
0 &=& \Big(\Ad(h') - 1 \Big) F(A) - \Ad(h') \Big( D(A) \delta\eta' 
- \delta\eta' \wedge  \delta\eta' \Big)  \nonumber \\ 
 & & + \delta F(\eta) + \delta\Big(\Ad(k) - 1 \Big) (B + F(\eta)) ~, \label{condi}
\eeqs
where the prime denotes the action of $\underline{\kappa}$ on
$\underline{\lambda}$. In the next section, we shall simplify
this condition. For this, however, we shall have to
equip our construction with some more structure.

\subsection{Consistency conditions}

There is a way to restrict fields in order to
make contact with the original hypercohomology. The idea is to
restrict the covariant derivatives on the various principal 
bundles so that they commute
with the \Cech-coboundary operator\footnote{
The formula (\ref{A-3}) is actually already an example of this: 
it just states that $\delta \Ad(h) = \Ad(g) \delta$. But the RHS is trivial
because the fields are on a torus.}.
In what follows we 
derive the relevant commutative 
diagrams adding some geometrical assumptions. Note, however,
that the rules of the previous chapter were not derived, but arouse 
as a natural extension of Abelian structure. The analysis below 
serves hence as a justification for these definitions.

The cocycle condition (\ref{B-A})
implies that $\delta B$ is a covariantly constant
section of the bundle where $A$ is the connection. This means 
that under $\underline{\lambda}$ for $\eta=0$ it transforms according to 
$\delta B'_{\alpha\beta} = \Ad(k^{-1}_{\alpha}h_{\alpha\beta}k_{\beta})\delta B_{\alpha\beta}$. 
On the other hand, we had already fixed $B$'s transformation properties in (\ref{B-k}). 
Hence $\Ad(h')\delta = \delta \Ad(k)$, which means that the diagram
\begin{equation}
\begin{CD}
\Omega^{[2]}_{\alpha} @>\Ad(k)>>    \Omega^{[2]}_{\alpha} \\
@VV{\delta}V   @VV{\delta}V \\
\Omega^{[2]}_{\alpha\beta} @>\Ad(h')>> \Omega^{[2]}_{\alpha\beta} 
\end{CD} \label{cd0}
\end{equation}
should commute. This assumption  relates the gauge transformations  
on twofold intersections so that $k_\alpha=k_\beta=h_{\alpha\beta}$.

In the Abelian case we found the globally defined three-form 
$H=\dd B_{\alpha}$ useful for distinguishing different gerbes. 
In the present situation we can build a covariant 
three-form under $\underline{\kappa}$ from $B_\alpha$ on ${\cal U}_\alpha$ 
by setting
\beqs
H_{\alpha} = D(\chi_\alpha) B_{\alpha}~. \label{Hdef}
\eeqs
The identity 
$ D(A) \delta B = \delta D(\chi) B$
would imply on twofold intersections that $H_{\alpha}$ 
extend to a section 
of the local bundle associated to $D(\chi_{\alpha})$. If this local bundle extends into
a global one, $H$ extends to its section. 
This compatibility constraint is natural in the sense that it is just 
the covariantization of the observation the the exterior derivative 
and the \Cech-coboundary operators commute in the diagram
\begin{equation}
\begin{CD}
\Omega^{[2]}_{\alpha} @>D(\chi)>>    \Omega^{[3]}_{\alpha} \\
@VV{\delta}V   @VV{\delta}V \\
\Omega^{[2]}_{\alpha\beta} @>D(A)>> \Omega^{[3]}_{\alpha\beta} 
\end{CD} \label{cd1}
\end{equation}
However, it is a restriction on $A_{\alpha\beta}$ and $\chi_\alpha$. 
The commutativity of the above diagram translates into the condition
\beqs
[A,\delta B] &=& \delta [\chi, B]~.
\eeqs 
We shall give later an explicit example.

If instead of acting on $B$, we consider the action on the one-forms $\eta$, the result
would be that the one-forms commute with each other $[\eta,\eta]=0$ and with $A$, namely
$[\eta, A]=0$. 
We shall be lead to this result presently, though through another route. 
The same argument puts the one-forms $\chi$ on the same torus on double intersections. 

Setting $h=k=1$ in (\ref{condi}) yields
\beqs
 F(A-\delta \eta) = F(A) - \delta F(\eta)~. \label{cond1}
\eeqs
We shall impose this formula as a restriction on $A$ and $\delta\eta$.  
This leads to $\delta F(\eta) = \dd \delta F(\eta)$
corresponding to the commutative diagram
\begin{equation}
\begin{CD}
\Omega^{[1]}_{\alpha} @>F(\eta)>>    \Omega^{[2]}_{\alpha} \\
@VV{\delta}V   @VV{\delta}V \\
\Omega^{[1]}_{\alpha\beta} @>\dd >> \Omega^{[2]}_{\alpha\beta} 
\end{CD} \label{cd2}
\end{equation}
One can verify that as was expected on general grounds, 
$[A,\delta \eta] = [\eta, \delta\eta] = 0$.  
Now (\ref{condi}) is identically satisfied. 
We shall have to ensure that the condition 
(\ref{cd1}) is consistent with
the transformation rules. This is not automatic, but we have 
to make yet a forth restriction 
\beqs
\delta D(\chi) F(\eta) =0~, \label{cond2}
\eeqs
or, equivalently, $\delta [\chi,F(\eta)]=0$.

In summary, we have had to assume the commutativity of diagrams 
(\ref{cd0}) and (\ref{cd1}), and that conditions (\ref{cond1}) and 
(\ref{cond2}) hold. All of these conditions are geometrical, and fit nicely 
together with Abelian hypercohomology.

\subsubsection*{A solution}

In order to see how these assumptions affect  
the differential forms it is useful to find 
concrete examples 
that satisfy them.
The geometrical picture that arises from 
these considerations restricts the various
fields in the following way:
\begin{itemize}
\item[{\bf (C1)}] The connections 
$\chi_\alpha \in \Omega^{[1]}({\cal U}_\alpha,\lie)$ 
define locally a subspace $ \ker\ad(\chi_\alpha) \in 
\Omega^{*}({\cal U}_\alpha,\lie)$. 
We can now choose the forms  $B_\alpha$ so that their  
restrictions on ${\cal U}_{\alpha\beta}$ belong there. 
Outside the double intersection there is no restriction.
\item[{\bf (C2)}]
Having hence fixed $\delta B$ on each  ${\cal U}_{\alpha\beta}$  
we have actually also fixed $F(A) = -\delta B$. Because $[A,\dd A] =0$ 
the connection $A$ and $\delta B$ should get their values in the same 
Cartan subalgebra.
\item[{\bf (C3)}]
On triple intersections  ${\cal U}_{\alpha\beta\gamma}$ 
this Cartan subalgebra should be a part of the algebra $\torus_{\alpha\beta\gamma}$ 
of a fixed torus  $T_{\alpha\beta\gamma}$.
\item[{\bf (C4)}] 
The \Cech{} 2-cocycle $g$ is built out of the transition functions $\phi$ 
of the local bundle $\underline{v}$ according to
\beqs
g_{\alpha\beta\gamma} =  
\phi_{\alpha\beta} \phi_{\beta\gamma} \phi_{\gamma\alpha} 
~\label{nong}~.
\eeqs
On ${\cal U}_{\alpha\beta\gamma}$
$g$ is constrained to lie in the fixed 
torus $T_{\alpha\beta\gamma}$. The torus could 
vary as one moves over the triple intersection.
\end{itemize}

Our construction is hence essentially Abelian
on triple and double intersections. The non-Abelianity 
of the construction lies outside the double intersections, 
and in the way in which these various locally 
Abelian constructions are related to each other. This means that the
restrictions appear rather as boundary conditions. The crucially non-Abelian
objects are the transition functions $\phi$, the one-forms $\chi$ and the two-forms $B$.

To see how this works, consider, for instance, how  
$\ker\ad(\chi_{\alpha})$ 
viewed as a collection of sections of the local principal bundle 
$\underline{v}$ transforms  
on a three-fold intersection ${\cal U}_{\alpha\beta\gamma}$ as we 
transport it around:
The transition functions of the local bundle $\underline{v}$ 
combine under this tour into $g$ as defined in (\ref{nong}).
This holonomy does not need to be trivial 
as we did not assume that the local bundles combine into a
global one. 
This far we did not impose much structure on $\underline{v}$. Let us now
suppose further that
$g_{\alpha\beta\gamma}$ happens to be an element of $T_{\alpha\beta\gamma}$ 
in order to satisfy condition C4 above. 
In addition, assume that on four fold intersections the $g$ 
are compatible in the sense that $\delta g = 0$. Despite 
the notation, (\ref{nong}) does not imply that $g$ would be 
exact (or even closed) 
as a collection of Abelian \Cech{} cocycles, because 
the transition functions $\phi$ are not in general Abelian. 

It turns out \cite{Finlay} that
the definition 
(\ref{nong}) of $g$
produces the right three-index object even if the transition functions 
$\phi$ are just general isomorphisms between principal bundles on different 
charts. Then the cocycle 
condition has to be modified, however, and  there 
does not seem to exist at 
present representations
of the underlying sheaves of grupoids in terms of 
differential geometry on them, or hypercohomology.

For concreteness, 
let us consider a toy model on a triple intersection ${\cal U}_{123}$, 
with the group $G=SU(2)$. 
Suppose $\chi_1 = x (\sigma^1 - \sigma^2)$, 
$\chi_2 = x (\sigma^1 + \sigma^2)$, and 
$\chi_3 = -x (\sigma^1 + \sigma^2)$. 
Then the transition functions can be taken constants 
$\phi_{12} = i \sigma^1$,  $\phi_{23} = i \sigma^3$, 
and  $\phi_{31} = i \sigma^2$. The resulting 
holonomy is $g_{ijk} = - \varepsilon_{ijk}$. 
The differences in the $B$ field are 
$\delta B_{12} = (B_2-B_1)\sigma^1  +(B_2+B_1)  \sigma^2$, 
$\delta B_{23} = - (B_3 + B_2) (\sigma^1 + \sigma^2)$, and 
$\delta B_{31} = (B_1 + B_3) \sigma^1 + (-B_1 + B_3)  \sigma^2$. 
In order to find the connections $A$ 
we have to assume $B_1 =0$. Then we can choose
one-forms $A_{ij} = \varepsilon_{ijk} A_k (\sigma^1 + \sigma^2)$. 
This also fixes the embedding of the 
torus $T_{ijk} \subset SU(2)$.

\subsubsection*{Transformations}

Given in the above sense consistent data $\underline{w}$, $\underline{v}$, let us now 
see what symmetries $\underline{\lambda}$, $\underline{\kappa}$ are left.
\begin{itemize}
\item[{\bf (S1)}]
On each patch $\underline{\kappa}$ acts as the gauge symmetries 
of the bundle $\underline{v}$.
\item[{\bf (S2)}]
On double intersections ${\cal U}_{\alpha\beta}$ these transformations are fixed to
coincide with the corresponding transformations of the gerbe $k_\alpha=h_{\alpha\beta}$.
\item[{\bf (S3)}]
On triple intersections also the transformations of the gerbe $h$ are fixed to respect the tori
$T_{\alpha\beta\gamma}$
\item[{\bf (S4)}]
The translations $\eta$ and their differences $\delta \eta$ 
belong on ${\cal U}_{\alpha\beta}$ to a Cartan subalgebra $\ker\ad(\chi_\alpha) \in 
\Omega^{*}({\cal U}_{\alpha\beta},\lie)$. 
\end{itemize}

In the previous toy model 
example the gauge transformations (S1) and (S2) make $\chi$ and $A$ 
into ordinary connections on the respective coordinate patches. 
The remaining shift symmetry $\eta$ can be found just as the $B_i$ were 
found above, except that now
$[\eta,\delta\eta]=0$. It follows, $\eta_1=0$, $\eta_2 = a(\sigma^1 + \sigma^2)$, and  
$\eta_3 = b(\sigma^1 + \sigma^2)$. It acts then on the triple intersections
just as in the Abelian case, along the fixed torus. Assuming $A_2=0$ we can also
choose $\eta_1 = a A_3 (\sigma^1 - \sigma^2)$ and $\eta_2=\eta_3=0$.

\subsection{Geometrical Interpretation}

The non-Abelian gerbe $\underline{w}$ found in the previous section
provides a tool to study the set of local, non-Abelian 
principal bundles $\underline{v}$. The local symmetry of the bundle
is frozen on double intersections so that gauge transformations
on both charts are identical. This transformation then acts
also on the gerbe $\underline{w}$. The two-form $B$ can be assumed
to commute with the connection $\chi$ on two-fold intersections. $B$
provides us an Abelian connection $A$ on double intersections. The translations 
$\eta$ are Abelian  on triple intersections  as well, and act on $A$ in the same way
as in the Abelian case. 

The gerbe $\underline{w}$ would then look exactly like $\rank~G$ copies 
of Abelian gerbes, were it not for the fact that $B$ is 
generally Lie algebra valued 
outside double intersections, and that $h$ can mix the diagonal elements of $A$ 
on double intersections. The crucial non-Abelianity resides 
in the principal
bundles $\underline{v}$, and $\underline{w}$ should be seen as 
an {\em almost} Abelian
obstruction for extending $\underline{v}$ into a global bundle. 

If the local bundles in  $\underline{v}$ are trivial, then its
sections can be conjugated to the Cartan subalgebrae fixed on 
various double intersections. The transition functions $\phi$
do still not have to be trivial, but they act as isomorphisms between these tori.
In particular, $g_{\alpha\beta\gamma}$ is an automorphism of the torus associated
to $\chi_\alpha$, and maps the torus back to itself thus permuting 
the diagonal elements. In this way, the gerbe can be used to describe
a {\em braid}.

\subsubsection*{Limitations}
\label{rajotus}

\Cech-cohomology does not depend on the choice of 
cover if the cover is fine enough \cite{Spanier}. However, in our 
discussions the cover is very particular. 
For instance, if there were enough 
three-fold intersections to cover the whole space 
the whole construction would collapse to $r$ copies
of Abelian gerbes. For our considerations it is, however, quite 
sufficient to know that there does exist a cover independent formulation
of non-Abelian gerbes \cite{Giraud, Finlay} that describes obstructions to
extend local bundles into global ones. We choose one of these configurations
together with a cover that is as simple as possible but still carries the
interesting information. In other words we smooth the system as 
much as possible, and try to push
the obstructions to trivializing it into as 
small and isolated neighbourhoods as possible.

We have found a very particular differential geometry description of these
objects as well. In that it seems to give an extension of Abelian hypercohomology 
the restrictions on the fields are under control.
One should ask, however, whether the parametric 
tori $T_{\alpha\beta\gamma}$ are actually necessary data. 
The observation that $g_{\alpha\beta\gamma}$ can become 
non-Abelian in general gives hope that
there actually might be a formulation where this data 
becomes superfluous. However, from the physical 
point of view the tori seem to be necessary, much as the 
physical Cartan subalgebrae in D-particle scattering, 
as we shall see presently.

\section{String backgrounds and gerbes}

In this section we make contact with the string 
theory considerations  of Section \ref{current},
where we coupled
the NS two-form fields and the Chan--Paton
vector fields to a non-Abelian current carried by a world-sheet. Three 
different symmetries acted on these fields {(G1)},  {(G2)}, and  {(G3)} 
in the notation of 
Section \ref{gaugesymm}. 
We also have an essentially Abelian gauge field from the RR sector, and a 
gauge symmetry associated to it. 

On the 
geometrical 
side there is
associated to the gerbe a local principal 
bundle $\underline{v} = [\phi,\chi]$. This should be 
identified with 
the Chan--Paton bundle on a D-brane. The 
obstruction to extend this bundle is $g$. 
The gauge symmetry  {(G3)} is then just the action of 
$\underline\kappa$. 

On two-fold intersections we have the essentially Abelian 
gauge field $A$. This should be the RR gauge 
field for the D-particle, or the D6-brane. 
The Chan--Paton gauge transformations 
were correlated to the RR gauge transformations $h$ on 
these two-fold intersections. 
If the action of these gauge transformations is not 
Abelian it seems that an isospin rotation
on the Chan--Paton sector induces a redefinition of 
which Cartan subalgebra
the RR fields live. 

We already noticed that the gauge 
transformations $h$ in $\underline{\lambda} = [h,\eta]$ 
connect the Chan--Paton
transformations and the RR gauge transformations. 
Also the transformations generated by $\eta$
play an important role. As they shift the $B$-field 
by the curvature $F(\eta)$ cf.~(\ref{B-eta}) they are
the natural generalization of the NS symmetry (G1).
The $\eta$ transformation also acts on the RR field in the way
NS transformation does. 
As was pointed out in \cite{Kalkkinen:1999uz} the Abelian 
version of the cocycle condition (\ref{B-A}) guarantees that  
the right gauge invariant field strength is
the same as in massive IIA supergravity, namely
\beqs
{\cal F}^{[2]} &=& F(A_{\alpha\beta}) + B_\beta
\eeqs
It then readily follows that the two-form $B$ in $\underline{w}$ is the
NS two-form.

The NS gauge invariant combination in open string 
theory $B + F(A_{\mathrm CP})$ appears here as well, but in the form  
${\cal F} = 
B_{\alpha} + F(\chi_\alpha)$. Its curvature is $H = D(\chi) {\cal F}$, 
as it should be, but $\underline{\lambda}$ does not seem to implement 
the NS symmetry (G1) correctly. Fortunately, all of the previous 
calculations on double intersections remain unchanged even if 
we extend the action of $\underline{\lambda}$ onto $\underline{v}$.
Then we have to assume 
again $[\chi,\eta]=[\eta,\eta]=0$, which 
makes $\eta$ into an effectively Abelian connection
so that $F(\chi-\eta) = F(\chi) - F(\eta)$, and ${\cal F}$ is again invariant. 

However, this would be pushing the NS symmetry too far. Though NS symmetry 
is present for Abelian currents coupled to the world-sheet -- 
which we have again correctly reproduced above in hypercohomology  --
it is not there for non-Abelian currents, as it heavily relies on 
the Abelianity of $F(A_{\mathrm CP})$. 
One should therefore think of the NS-symmetry $(G1)$ rather 
as a freedom to redefine
the connection $\chi$ by shifting it with suitably Abelian 
form. Our construction therefore necessitates a nontrivial
non-Abelian extension of the NS-symmetry. 
There is exactly the same interplay 
between Abelian and non-Abelian currents in the effective supergravity 
Lagrangians and the above generalized hypercohomology. 

Let us finally consider conserved charges. The local bundles 
$\underline{v}$ are classified
by the Chern class ${\mathrm{Ch}}(F(\chi_\alpha))$. The bundles in $\underline{w}$ 
also have nontrivial first Chern class 
$\mathrm{ch}_1(F(A_{\alpha\beta}))$. The invariant quantity 
associated to $B$ is $\tr ~H = \tr ~D(\chi)B$.
Consider its integral over a sphere $S^3$ that is divided into 
two discs ${\cal U}_\alpha$, 
${\cal U}_\beta$, whose boundaries $S^2$ coincide. Then
\beqs
Q_{{\mathrm NS}} = \int_{S^3} \tr ~H = \int_{S^2} \tr ~(B_\alpha - 
B_\beta) = \int_{S^2} \tr ~F(A_{\alpha\beta}) ~. \label{varaus}
\eeqs
Thus NS charge is non-trivial, if $\tr F(A_{\alpha\beta}) $ has 
monopole number, i.e.~there are D6-branes \cite{Kalkkinen:1999uz}. 
The NS charge is well defined under the $\eta$ shifts as well, because
\beqs
\int_{S^2} \tr~\delta F(\eta_{\alpha\beta}) = \int_{S^2} 
\tr~\dd~\delta\eta_{\alpha\beta} =0 ~.
\eeqs
For fixed bases of RR fields these formulae yield 
charges that do not depend on $\eta$ or the choice of
homology cycles, even if the traces are dropped in (\ref{varaus}).

\section{Conclusions}

We started by studying a branched cover of space-time and showed
how the NS two-form fields are made to carry the same Lie algebra indices
that the Chan--Paton gauge fields have. 
Much in the same way that the latter fields are promoted to non-Abelian 
Lie algebra fields in the case of a stack of
coinciding D-branes, we argued that there should appear additional light 
degrees of freedom from strings that connect D-branes on different branches
of the space-time. A DBI action argument was also used to indicate which 
symmetries there should be present.

The curvature of the NS $B$ field appears 
on the level of effective supergravity as the 
characteristic class of a gerbe. 
In order to set the stage for addressing dynamical 
issues concerning this non-Abelian
$B$ field it is therefore necessary to 
generalize the Abelian hypercohomology 
construction. This we did, and the resulting structure
incorporates strikingly well, and in particular 
without introducing unphysical degrees of freedom, 
all the relevant supergravity fields and symmetries.

This construction sheds light on the difficulties encountered
in trying to describe perturbatively for instance the exotic $N=(0,2)$
theories in six dimensions. In the case of non-Abelian Yang--Mills the 
right object to study in supergravity seems to be the Wilson line, 
i.e.~the holonomies of the principal bundle. It seems therefore that  
the right strategy to attack the dynamical problem here should be, analogously, to 
understand the holonomies of the gerbe using the techniques developed here. 
These very same holonomies arise also in guaranteeing that the string world-sheet
measure is anomaly free. For instance
the analysis in \cite{Freed:1999vc} 
was concerned in essentially defining
the the holonomy of an Abelian gerbe.

\vspace{1cm}
\noindent
{\bf Acknowledgements}: I thank L.~Bonora, R.~Iengo, and F.~Thompson 
for useful discussions, and in particular 
P.~Tran-Ngoc-Bich for collaboration in the early stages of this project. 
This work was supported in part by the European Union TMR program CT960045.

\end{document}